# Molecular Dynamics Simulations of the Johari-Goldstein Relaxation in a Molecular Liquid

D. Fragiadakis and C.M. Roland

Naval Research Laboratory, Chemistry Division, Code 6120, Washington DC 20375-5342

ABSTRACT: Molecular dynamics simulations (mds) were carried out to investigate the reorientational motion of a rigid (fixed bond length), asymmetric diatomic molecule in the liquid and glassy states. In the latter the molecule reorients via large-angle jumps, which we identify with the Johari-Goldstein (JG) dynamics. This relaxation process has a broad distribution of relaxation times, and at least deeply in the glass state, the mobility of a given molecule remains fixed over time; that is, there is no dynamic exchange among molecules. Interestingly, the JG relaxation time for a molecule does not depend on the local density, although the non-ergodicity factor is weakly correlated with the packing efficiency of neighboring molecules. In the liquid state the frequency of the JG process increases significantly, eventually subsuming the slower $\alpha$-relaxation. This evolution of the JG-motion into structural relaxation underlies the correlation of many properties of the JG- and $\alpha$-dynamics.

---

In addition to the structural ($\alpha$-) relaxation, glass-forming substances almost universally show a faster process called the Johari-Goldstein (JG) relaxation. Evident in the mechanical, electrical, and thermal properties of materials, this relaxation involves reorientation of all atoms in the molecule, and thus is present even in rigid chemical structures [1,2]. This differentiates the JG process from secondary dynamics involving intramolecular degrees of freedom, such as motion of pendant groups. JG motion is usually observed below the glass transition temperature, often dominating the relaxation behavior of the glass. At higher temperatures it either merges with the $\alpha$-relaxation (*merging* scenario) or the intensity of the $\alpha$-relaxation goes to zero as the JG dynamics evolves into the structural relaxation (*splitting* scenario) [3]. In either case, the JG process is intimately related to structural relaxation. Other evidence for its connection to the glass transition includes a change of the temperature-dependence of both the JG-relaxation time, $\tau_{JG}$, and its relaxation strength as $T_g$ is traversed (although non-JG secondary relaxations sometimes show similar behavior) [4-7], and correlations of $\tau_{JG}$ and its activation energy with either the non-exponentiality or the fragility of the α-process [8-12]. Experimental results indicating that the JG-relaxation senses the thermodynamic variables underlying the glass transition also show that it serves as the precursor to structural relaxation [2].

Although the significance of the JG dynamics is well appreciated, important aspects of the process remain unclear. Historically there have been two distinctly different hypotheses for why reorientation of a molecule can occur in the glassy state. One envisages islands of mobility, due to loose packing, that enable some molecules to undergo motions precluded generally by the frozen structure of the glass [13]. An alternative view is that virtually all molecules participate in the JG relaxation via small-angle rotations, prior to larger-angle reorientations associated with structural relaxation [14,15]. NMR and solvation dynamics experiments have shown that at least near $T_g$, most molecules undergo the JG relaxation [14,15]; however, this result is not easily reconciled with the temperature- and aging-dependences of the JG relaxation strength [16]. The mechanism for the JG process affects the interpretation of the highly non-exponential nature of the relaxation, which is observed even though the temperature dependence of $\tau_{JG}$ in the glass is Arrhenius. Is this non-exponentiality inherent or due to ensemble averaging of exponential decays with spatially varying time constants? Even the basic molecular motions responsible for the JG process are far from clear. NMR measurements on simple organic glass formers suggest angular jumps of a few degrees, independent of temperature; however, in other materials the JG relaxation comprises large-angle reorientations. For example, in polymethylmethacrylate the pendant group undergoes 180° flips that are coupled to rocking motion of the chain backbone [17,18].

To address these issues we employed molecular dynamic simulations (mds) to study the JG relaxation in a simple molecular glass-former. Existing mds of the JG process have been limited to polymers: Bedrov and Smith investigated the JG process in polybutadiene [19] and in a bead-chain model polymer [20]. Simulations of supercooled mixtures of soft spheres or simple molecular liquids typically show only a two-step relaxation, consisting of vibrations at short times and a longer time α-relaxation, with no JG process apparent in the studied time scales. Hints of the JG process may be found in mds of diatomic molecular liquids. In symmetric [21] or weakly asymmetric [22-24] rigid diatoms having short lengths, 180° flips become prominent in the rotational dynamics. These reorientations enable the odd orientational degrees of freedom (referring to the parity of the Legendre polynomial describing the orientation of the molecule) to completely relax, with the relaxation time having an Arrhenius temperature dependence, even in the translationally arrested glassy state; the even degrees of freedom, however, remain frozen. While such behavior has characteristics reminiscent of a secondary relaxation, it differs from experimental observations in actual glass-forming substances, where both the first order (measured by dielectric spectroscopy) and second order (measured by NMR or dynamic light scattering) rotational correlation functions only partially relax via the JG process, decaying to a finite, usually large, value in the glass. Herein we carry out mds of rigid (fixed bond length) diatomic molecules having a larger degree of asymmetry, and observe dynamics which we identify with the experimentally observed JG relaxation in real glass-forming materials.

The simulations were carried out using the HOOMD simulation package [25,26]. The system studied was a binary mixture (800:200) of asymmetric diatomic molecules labeled AB and CD. Atoms belonging to different molecules interact through the Lennard-Jones potential

$$U(r) = 4\varepsilon_{ij}\left[\left(\frac{\sigma_{ij}}{r}\right)^{12} - \left(\frac{\sigma_{ij}}{r}\right)^{6}\right]$$

where $r$ is the distance between particles, and $i$ and $j$ refer to the particle types A, B, C and D. The energy and length parameters $\varepsilon_{ij}$ and $\sigma_{ij}$ are chosen based on the Kob-Andersen (KA) liquid, a mixture that does not easily crystallize [27]. This was done as follows (noting that alternative choices of $\varepsilon_{ij}$ and $\sigma_{ij}$ gave qualitatively the same results): The energy parameters $\varepsilon_{ij}$ are those of the K-A liquid; i.e. $\varepsilon_{AA} = \varepsilon_{AB} = \varepsilon_{BB} = 1.0$, $\varepsilon_{CC} = \varepsilon_{CD} = \varepsilon_{DD} = 1.0$ and $\varepsilon_{AC} = \varepsilon_{AD} = \varepsilon_{BC} = \varepsilon_{BD} = 1.5$. To set $\sigma_{ij}$, we use the original K-A parameters for the larger A and C particles, while the smaller B and D particles have a size 65% that of A and C, respectively. Therefore, $\sigma_{AA} = 1$, $\sigma_{CC} = 0.88$, $\sigma_{BB} = 0.65$, $\sigma_{DD} = 0.65 \times 0.88$. For the interactions between different types of particles, we take $\sigma_{ij} = S_{ij}(\sigma_{ij} + \sigma_{ij})$ where $S_{ij}$=0.5 (additive interaction) when the particles are the same type ($i,j$ = AB, CD), and $S_{ij} = 0.4255$ when the particles belong to different types ($i,j \neq$ AB, CD), the latter chosen to give the KA value for $\sigma_{AC} = 0.8$. All atoms have a mass $m$=1. The bond lengths A-B and C-D were fixed to 0.4 using rigid body dynamics [28]. All quantities are expressed in units of length $\sigma_{AA}$, temperature $\varepsilon_{AA}/k_B$, and time $(m\sigma_{AA}^2/\varepsilon_{AA})^{1/2}$.

Simulations were carried out in an $NPT$ ensemble at a constant pressure $P=1$, at 16 temperatures between $T$=0.25 and $T$=1.5. The time step was 0.005 for higher $T$ and 0.01 for lower $T$. Data were collected at each temperature after an equilibration run several times longer than the structural relaxation time $\tau_\alpha$. At low temperatures ($T < 0.56$), structural relaxation is extremely slow, and translational and orientational correlation functions do not decay to zero over the duration of the simulation runs; i.e., the system is out of equilibrium. For these conditions we increased the equilibration runs (to ~$10^7$ steps), until neither significant drift in volume nor aging of the translational and rotational correlation functions were observed; the residual rotational motion of the molecules at these temperatures takes place within a non-equilibrium, essentially frozen structure.

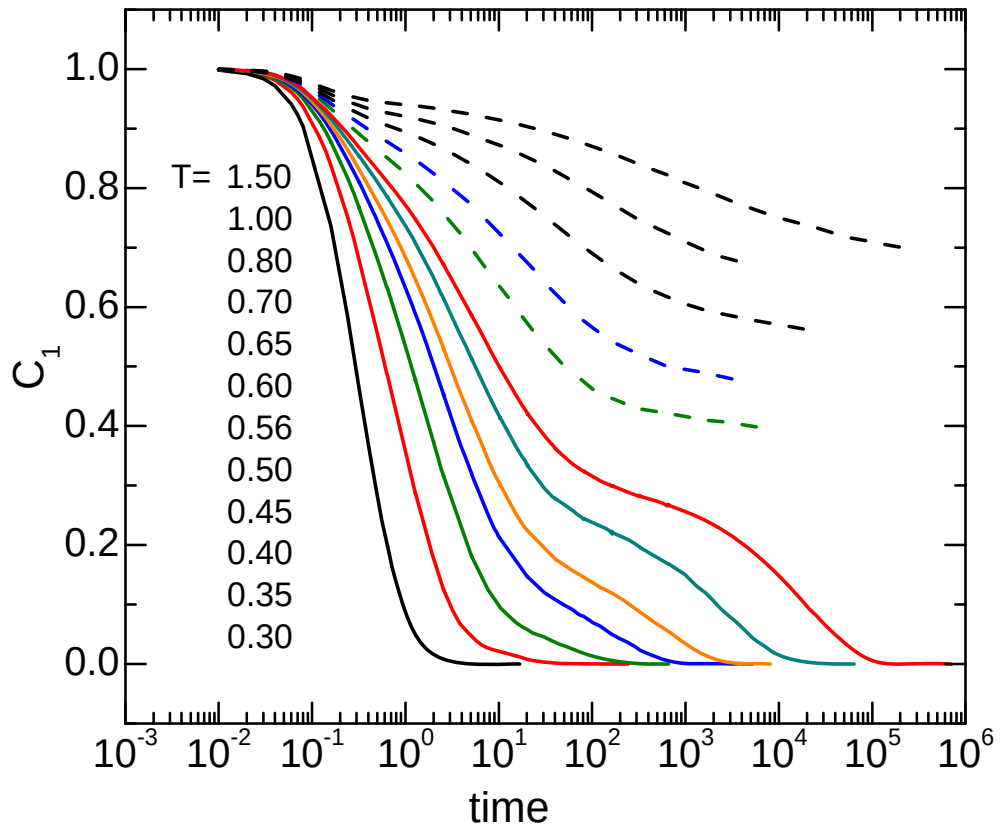

Fig. 1. First-order rotational correlation function for the AB molecules at P=1 and the indicated temperatures.

Figure 1 shows for various temperatures the first-order orientational correlation function for the AB molecules, $C_1(t) = \langle cos\,\theta(t)\rangle$ where $\theta$ is the angle between the vector AB at times 0 and $t$. Unlike the case of symmetric diatomic molecules, higher order correlation functions (not shown) have qualitatively similar behavior. At all temperatures, a small decrease in $C_1$ due to oscillations within the local structure formed by neighboring molecules (cage rattling) takes place at a temperature-independent $t \cong 0.1$. At high temperatures, $C_1$ then decays to zero via stretched-exponential structural relaxation. Below a temperature $T_{on} \cong 1.0$, the relaxational component splits into the shorter-time secondary and longer-time α-processes; the latter appears at $T_{on}$ as a long-time tail, which grows in intensity with decreasing temperature. Translational diffusion of the molecules, as well as structural relaxation (e.g., evolution of the volume after a step change in temperature) exhibits a temperature-dependence similar to that of $\tau_\alpha$. We define the glass transition temperature as $\tau_\alpha(T_g) = 10^6$, which is the typical time for the longest simulation runs, obtaining $T_g = 0.52$. Below $T_g$, the system is in the glassy state: the molecules do not translate and the system is out of equilibrium. However, the rotational correlation function $C_1$ significantly relaxes via secondary motions, reaching a plateau at a nonzero value of the non-ergodicity parameter, $Q$. The magnitude of this plateau increases with decreasing temperature. Because of the rigid nature of the molecule, the only local motions available in the glassy state involve the entire molecule; therefore, the dynamics observed in the glassy state corresponds to the JG process.

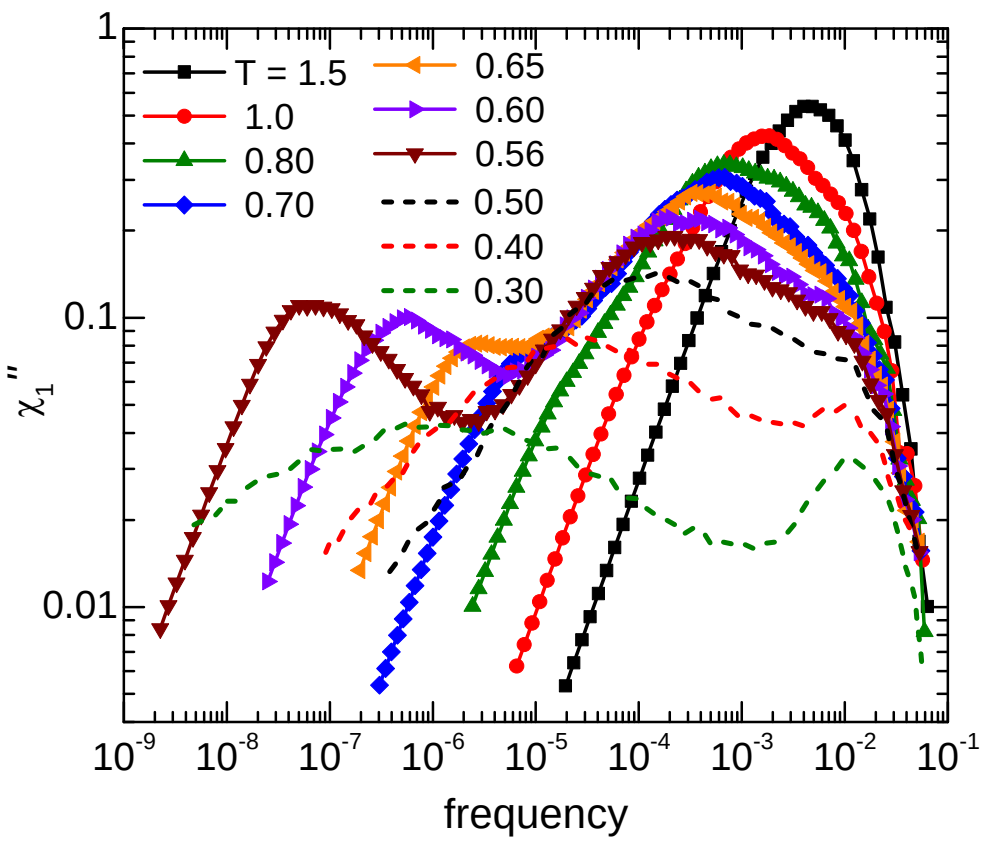

Fig. 2. Imaginary part of the susceptibility associated with the first-order rotational correlation function of the AB molecules.

The above results may be compared directly with experimental data (such as dielectric and light scattering spectra) by examining the imaginary part of the susceptibility, calculated from the Fourier transform (FT) of the $C_1(t)$ in Fig. 1 and displayed in Figure 2. The cage rattling contribution is more evident at lower temperatures, at a temperature-independent frequency of *ca.* $10^{-2}$. The JG-process ensues at lower frequencies. With decreasing temperature, the slower α-relaxation separates from this secondary process, eventually below $T_g$ falling outside the accessible frequency window. The α-process is relatively narrow, and is well described by the FT of a stretched exponential function, with an exponent ~ 0.8 close to $T_g$ and increasing at higher temperatures. The JG relaxation is broad, increasingly so on cooling, and symmetric; it can be accurately fit by a Cole-Cole function.

Figure 3 displays the variation with temperature of the relaxation times and strengths for both processes. The JG relaxation has the expected Arrhenius behavior in the glassy state, while above $T_g$ some curvature in log $\tau_{JG}$ vs. $1/T$ plots is evident. The JG relaxation strength increases with increasing temperature, but that of the $\alpha$-process shows the opposite behavior, going to zero at the onset temperature $T_{\mathrm{on}}$. These trends reflect the behavior observed experimentally in the dielectric strength and relaxation times of supercooled liquids, in particular those exhibiting the "splitting scenario" for the α-β crossover region such as polymethylmethacrylate [3]. Also plotted in Fig. 3 is the translational diffusion coefficient $D_{\mathrm{AB}}$ of the AB molecules, calculated from the long-time behavior of the mean square displacement of the molecular center of mass. $D_{\mathrm{AB}}$ and $\tau_\alpha$ have similar temperature dependences, which confirms that the $\alpha$-relaxation observed in the rotational motion is coupled to translation of the molecules.

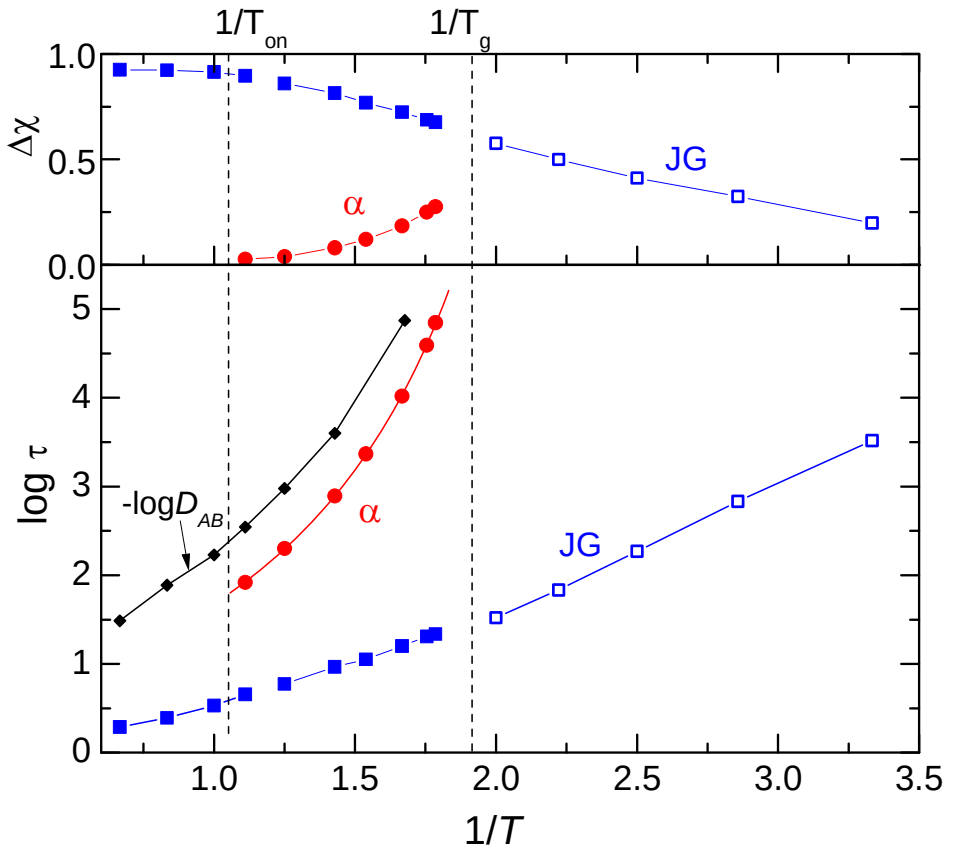

Fig. 3. Temperature dependence of the α- and JG-relaxation times and intensities. The temperatures associated with the respective appearance and disappearance of the α-process are indicated.

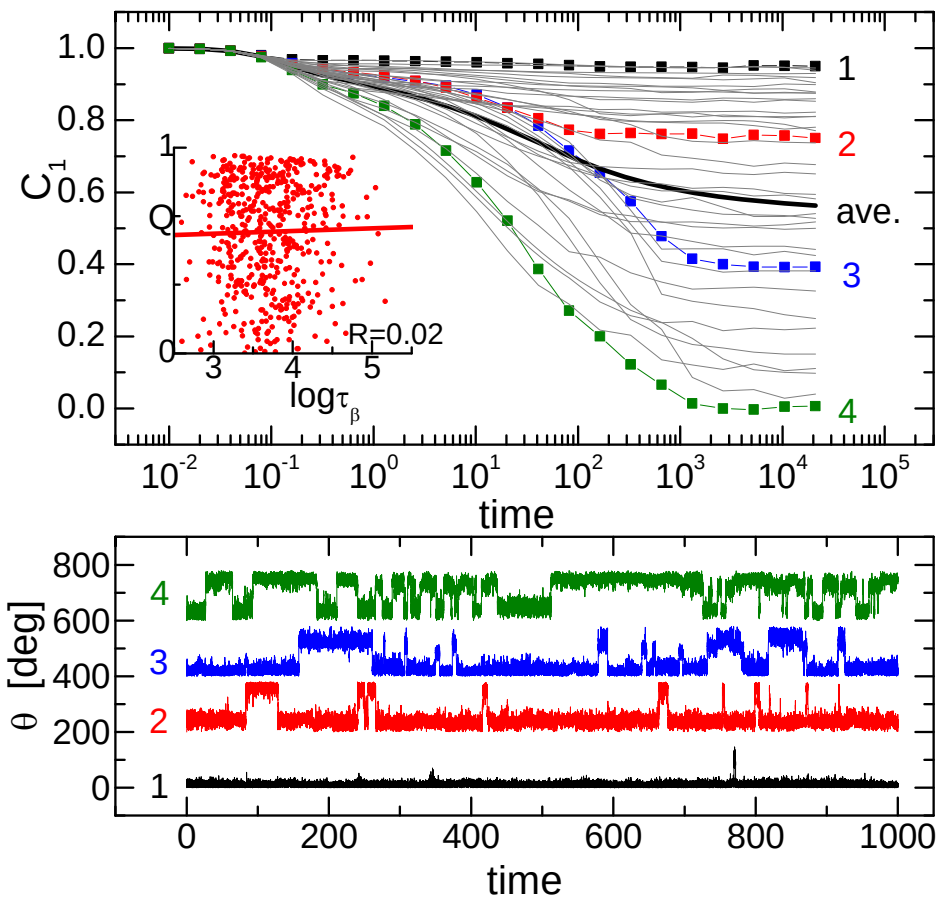

Fig. 4. (top) Rotational correlation function for representative individual AB molecules (gray lines). Inset: single-molecule non-ergodicity factor Q against relaxation time $\tau_{JG}$, along with the linear regression and correlation coefficient. (bottom) Angular position as a function of time for the four molecules denoted by the symbols in the upper panel. The curves have been shifted vertically for clarity.

We focus next only on the glassy state, $T < T_g$. In this regime we are able to observe the system for times much longer than $\tau_{JG,}$ but much shorter than the structural relaxation time. We examine the contribution of individual molecules to the relaxation behavior to directly assess heterogeneity of the JG relaxation in the glass. In the upper panel of Figure 4 (thin lines) are the decay curves for representative molecules at $T$ = 0.4. There is a broad distribution of both the time constants and the amount of relaxation (plateau value $Q$) for individual molecules. The relaxation for each is significantly narrower than the average process, but broader than a Debye relaxation. The inset is a plot of log $\tau_{JG}$ versus $Q$ for one temperature, with each data point representing an individual molecule. Unlike for the average over all particles, there is no correlation between the rapidity of the decay of the orientational correlation function of a single molecule and the magnitude of this decay. Over the time scale of the JG-relaxation, there is no mobility exchange among particles; that is, the more mobile particles remain mobile and likewise for the immobile species (and this is true whether mobility is defined by $Q$ or $\tau_{JG}$). The behavior does change even as $T_g$ is approached, since dynamical exchange requires motion over length scales exceeding that associated with the JG process alone.

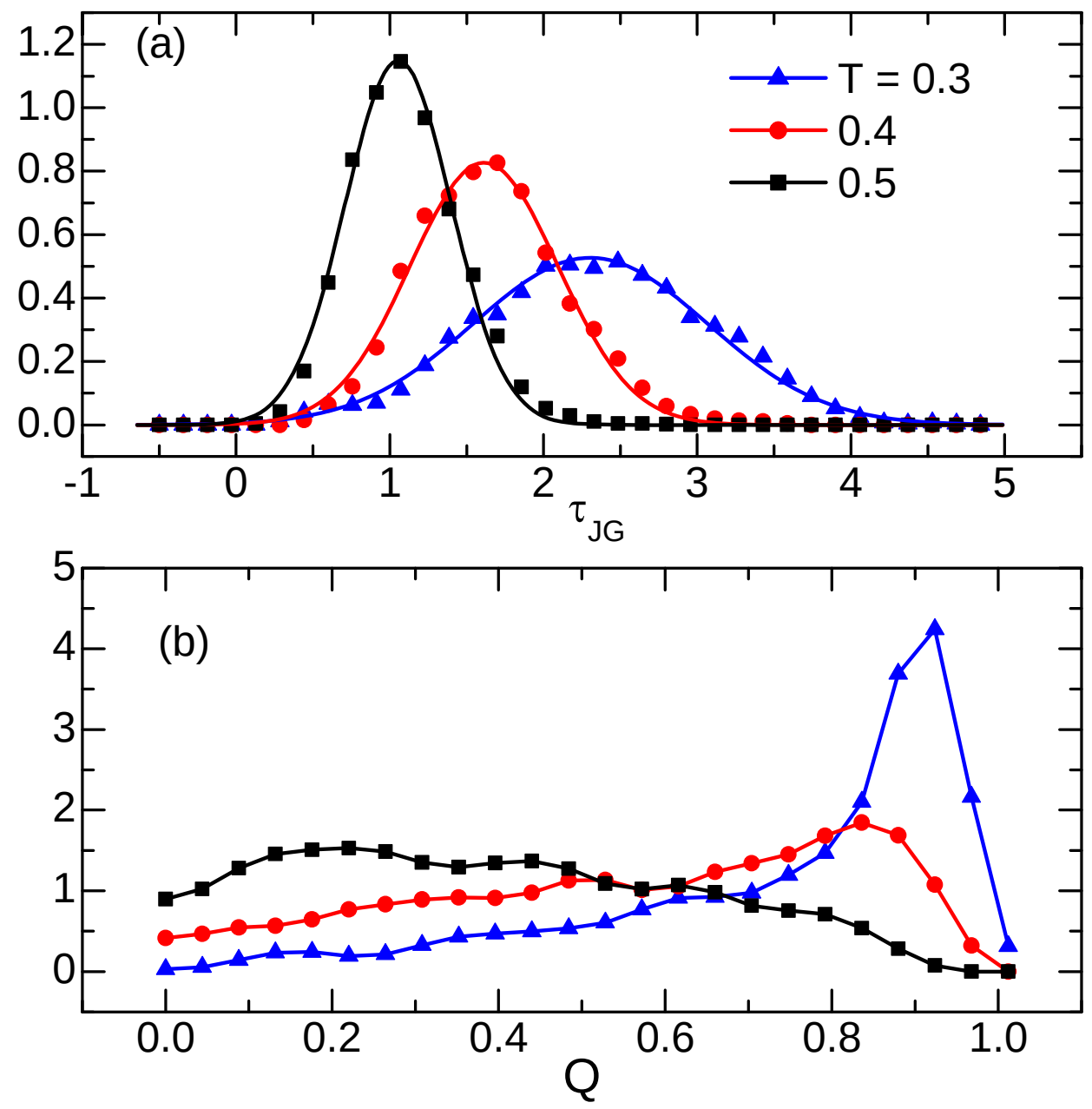

Fig. 5. Distribution of: (a) single-molecule JG-relaxation times and (b) single-molecule non-ergodicity parameters of the AB molecules, for the indicated temperatures in the glassy state.

For the four individual molecules denoted by symbols in the upper panel in Fig. 5, we show in the lower panel their orientation as a function of time. Molecular motion in the glassy state is characterized by rather large angular jumps superimposed on the rapid oscillatory motion. These changes in orientation correspond to transitions between discrete minima in the potential energy surface. The frequency of the angular jumps governs the JG relaxation time. The symmetry of the jumps (probability that the orientation will be reversed within a relatively short time) governs the amount of relaxation (non-ergodicity factor); that is, the intensity contributed to the JG relaxation by a given molecule. Extrapolating to above $T_g$, we can ascribe the decreasing intensity of the $\alpha$-relaxation with temperature to the diffusive motions becoming subsumed by the molecular flips occurring on the $\tau_{JG}$ timescale.

Figure 5a shows the distributions of JG relaxation times for individual molecules at each of four temperatures. Because of the distribution, the relaxation function will always be non-exponential, regardless of the shape of the decay for individual species. The distribution of $\tau_{JG}$ is broad and symmetric, well described by a Gaussian function, and broadens with decreasing temperature. The corresponding distributions of the non-ergodicity factor are shown in Figure 5b. Close to $T_g$ there is a very broad distribution of $Q$. With decreasing temperature, a large fraction of molecules has $Q > 0.9$, each of these contributing very little to the JG relaxation strength. This broad, temperature-dependent distribution embodies aspects of both the "islands of mobility" and homogeneous relaxation scenarios, commonly discussed for the JG process.

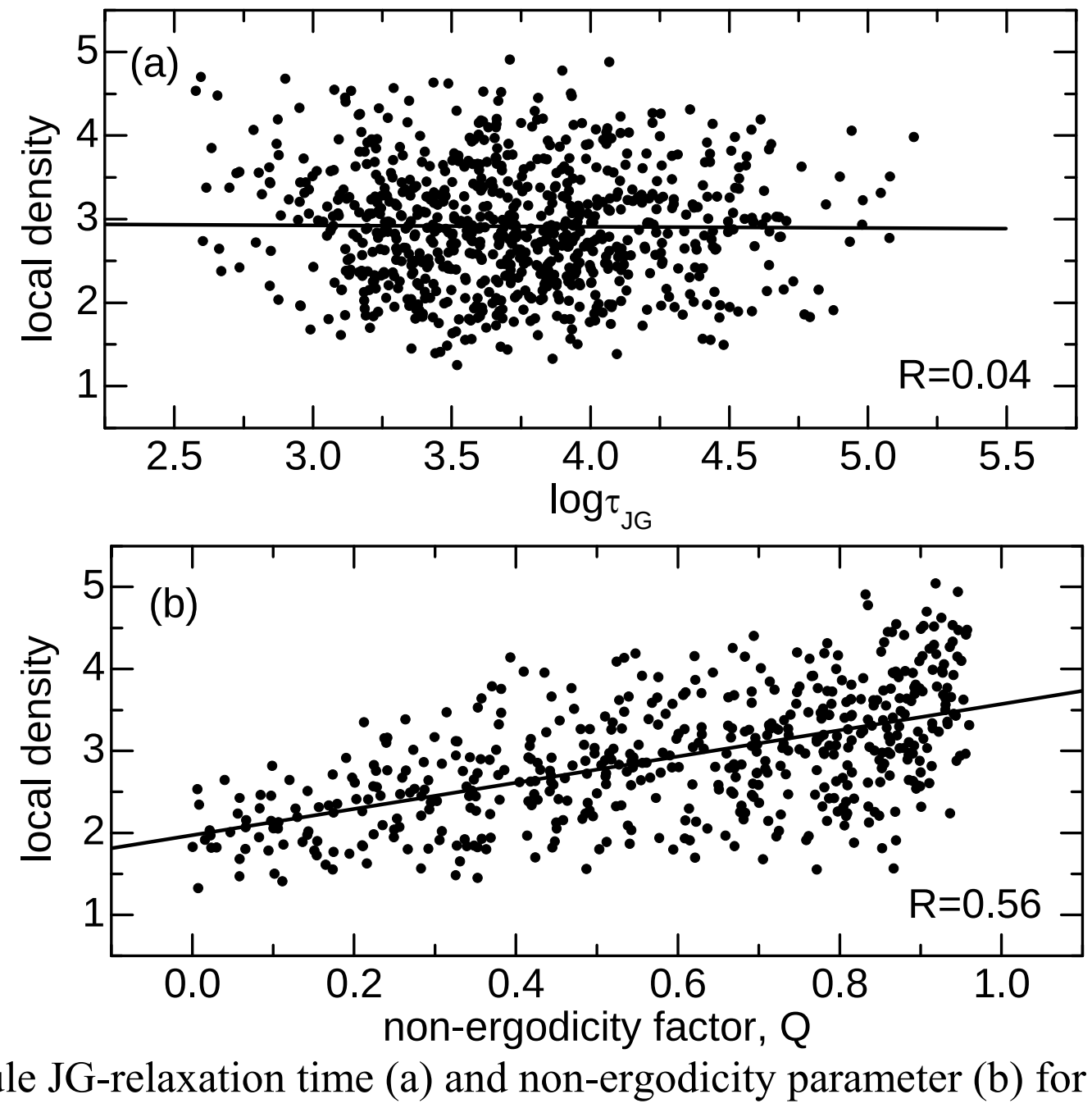

Fig. 6. Single molecule JG-relaxation time (a) and non-ergodicity parameter (b) for the AB molecules plotted against local density (density in a sphere of radius 0.8 around the B particle). The lines represent linear regressions having the indicated correlation coefficients.

To further examine mechanisms for the JG dynamics, we investigated the relationship of the local structure to relaxation in the glass. The local density around each particle was determined by enumerating the number of particle centers within a sphere of radius equal to 0.8 (twice the bond length of our simulated diatomic molecule) around the smaller atom of each molecule. We find there is no correlation of this local density with the relaxation rate of the particle (Figure 6a). The JG-relaxation time of individual molecules is determined by factors beyond just the local packing. On the other hand, there is a modest correlation between the non-ergodicity factor and the local density (Figure 6b); the linear correlation coefficient between the two quantities is 0.56. These results are consistent with the absence of correlation between the decay rate and the extent of the relaxation for single molecules (Fig. 4a inset).

In summary, we have observed the Johari-Goldstein dynamics in asymmetric diatomic molecules in the liquid and glassy states. In the glass the JG relaxation consists of large-angle rotational jumps. At temperatures above $T_g$, these large-angle reorientations increase their frequency and eventually dominate the $\alpha$-relaxation; thus, the JG-motion in the glass evolves into the structural relaxation of the liquid. It is for this reason that the properties of the JG and $\alpha$ relaxations are correlated. Concerning the putative dichotomy between the two mechanisms proposed for the JG process, islands of mobility versus all molecules participating with dynamic exchange, both interpretations are supported to some degree by the mds results, with their relative contribution changing with proximity to the glass transition. There is a weak correlation of the single-molecule non-ergodicity factor with the local density; however, the packing efficiency has no direct effect on the magnitude of the reorientation rate of individual molecules in the glassy state. This decoupling of $\tau_{JG}$ and Q follows from the nature of the potential energy surface in the glass: the probability of a molecule being trapped in a non-equilibrium configuration is unrelated to the rapidity of the local reorientations among shallow minima of the potential.

This work was supported by the Office of Naval Research, in part by Code 331.

REFERENCES

[1] G.P. Johari and M. Goldstein, J. Chem. Phys. **53**, 2372 (1970).

[2] K.L. Ngai and M. Paluch, J. Chem. Phys. **120**, 857 (2004).

[3] E. Donth, *The Glass Transition: Relaxation Dynamics in Liquids and Disordered Materials*, Springer Series in Materials Science Vol. 48, (Springer, Berlin 1998).

[4] H. Wagner and R. Richert, J. Phys. Chem. B **103**, 4071 (1999).

[5] T. Fujima, H. Frusawa and K. Ito, Phys. Rev. E **66**, 031503 (2002).

[6] R. Nozaki, H. Zenitani, A. Minoguchi, and K. Kitai, J. Non-Cryst. Solids **307–310**, 349 (2002).

[7] M. Paluch, C. M. Roland, S. Pawlus, J. Ziolo and K.L. Ngai, Phys. Rev. Lett. **91**, 115701 (2003).

[8] K.L. Ngai, J. Chem. Phys. **109**, 6982 (1998).

[9] A. Kudlik, S. Benkhof, T. Blochowicz, C. Tschirwitz and E. Roessler, J. Molec. Struc. **479**, 201 (1999).

[10] S. Capaccioli, D. Prevosto, K. Kessairi, M. Lucchesi, P. Rolla, J. Non-Cryst. Solids **353**, 3984 (2007).

[11] K.L. Ngai and S. Capaccioli, Phys. Rev. E **69**, 031501 (2004).

[12] R.B. Bogoslovov, T.E. Hogan and C.M. Roland, Macromolecules **43**, 2904 (2010).

[13] G.P. Johari and M. Goldstein, J. Chem. Phys. **53**, 2372 (1970).

[14] H. Wagner and R. Richert, J. Non-Cryst. Solids **24**, 19 (1998)

[15] M. Vogel and E. Roessler, J. Chem. Phys. **114**, 5802 (2000).

[16] J.K. Vij and G. Power, J. Non-Cryst. Solids **357**, 783 (2011).

[17] K. Schmidt-Rohr, A. S. Kulik, H. W. Beckham, A. Ohlemacher, U. Pawelzik, C. Boeffel, and H.W. Spiess, Macromolecules **27**, 4733 (1994).

[18] S.C. Kuebler, D. J. Schaefer, C. Boeffel, U. Pawelzik, and H.W. Spiess, Macromolecules **30**, 6597 (1997).

[19] D. Bedrov and G.D. Smith, Phys. Rev. E **71**, 050801, (2005).

[20] D. Bedrov and G.D. Smith, J. Non-Cryst. Solids **357**, 258 (2011).

[21] A.J. Moreno, S.-H. Chong, W. Kob, and F. Sciortino, J. Chem. Phys. **123**, 204505 (2005).

[22] C. De Michele and D. Leporini, Phys. Rev. E **63**, 036702 (2001).

[23] S. Kaemmerer, W. Kob and R. Schilling, Phys. Rev. E **56**, 5450 (1997)

[24] M. Higuchi, J. Matsui and T. Odagaki, J. Phys. Soc. Japan **72**, 178 (2003).

[25] HOOMD web page: http://codeblue.umich.edu/hoomd-blue

[26] J.A. Anderson, C.D. Lorenz and A. Travesset, J. Comp. Phys. **227**, 5342 (2008).

[27] W. Kob and H.C. Andersen, Phys. Rev. E **48**, 4364 (1993).

[28] T.D. Nguyen, C.L. Phillips, J.A. Anderson, S.C. Glotzer Comp. Phys. Comm. **182**, 2307 (2011).